# Testing machine for Expansive Mortar

R. A. V. Silva [1]

[1]Departamento de Engenharia de Materiais – Universidade Federal de Campina Grande – PB

___

Abstract:

The correct evaluation of a material property is fundamental to, on their application; they met all expectations that were designed for. In development of an expansive cement for ornamental rocks purpose, was denoted the absence of methodologies and equipments to evaluate the expansive pressure and temperature of expansive cement during their expansive process, having that data collected in a static state of the specimen. In that paper, is described equipment designed for evaluation of pressure and temperature of expansive cements applied to ornamental rocks.
.

___

## 1. Introduction

A mortar is a material resulting from a mixture of aggregates (fine), one or more binders, water and any additives in order to improve their properties. The mortars are designed to meet the most common functions such as settlement and the coat of masonry elements (walls, columns, facades, etc.).. Besides these commonly used mortars, we still have some special-purpose, among which stands out in this paper expansive mortar.

The mortar is an expansive non-explosive demolition agent, powder, whose major component is the quicklime. In contact with water, begin hydration reactions, with an increase in volume during the progress of these reactions, promoting, when in confinement, great pressure on the confining walls, which reach approximately 78 MPa. These stresses cause cracks in the confining medium (rock, concrete or other means you want to demolish), whose magnitude and direction depend on the balance of active efforts in that medium.

The parameters of the current manufacturers that is taken as reference is that the beginning of the reaction must occur 15 to 30 minutes after the addition of water pressure and has expansive than ton/m2 7000, a period of up to 30 hours after application in the hole (CAIMEX, 2010).

These manufacturers typically have different formulations for each temperature range, the main variables that are modified in these formulations are takes time and evolution of forces in the middle. Another relevant factor is that the expansion should take place without overflow hole, ie, following an order to preferentially unilateral expansion, which is obtained by hardening of the mortar in contact with the expanding air in the orifice area quickly after application.

Huynh & LAEFER, (2009) cite the composition of expansive mortar shown in Table 1.

Table 1: Chemical composition of expansive mortars

| Substance | Fraction (%) |
|---|---|
| $SiO_2$ | 1,5-8 |
| $Al_2O_3$ | 0,3-0,5 |
| $Fe_2O_3$ | 0,2-3 |
| CaO | 81-96 |
| MgO | 0,0-1,6 |
| $SO_3$ | 0,6-4 |

In literature, there are applications of tests commonly used in materials characterization of materials, such as electron microscopy, x-ray, DTA, TGA and others. The tests for the validation of the product are found in the dismantling of direct application in confining means for, after some time, check out the action during the expansion process. These tests were performed, mostly just so visual, and even when there were quantitative assessments, these were performed on a non-reproducible, as in HANIF & AL-Maghrabi (2006) and ISHIDA (2005).

Shiro Ishi (2006) reports that business is an expansive mortar composed of lime, clay and plaster mixed in certain proportions. This mixture is calcined in a rotary kiln 1500oC. The resulting mass is then ground until 2000 to 3000cm2 / g specific area of the grains.

According to Kawano et al. (1982) these mortars are prepared by spraying a clinker obtained mainly by mixing calcium oxide, silicon oxide (SiO2) and calcium sulfate (CaSO4). According to Moyer et al. (1980) is also formed a mixture of clinker, which he highlights the use of calcium oxide (80 - 95%), portland cement, calcium hydroxide (Ca (OH) 2) and calcium carbonate (CaCO3).

In this paper, we propose a methodology and equipment where they collect the relevant data on the behavior of expansive mortar, and this data was collected from a platform on which the specimen containing the material is subjected to a static state, in conditions similar to its application in the field, but reproducible. The parameters obtained through a standardized methodology are needed to control the quality of the mortar developed, especially in view of comparative data obtained in similar tests with mortar business.

**2.1 Testing Machine**

To develop this machine, followed by two lines of research. The first involves the mechanical part shown in Figure 1, comprising a structure that requires a body of evidence including a static state specimen and media assistants, and the second comprising the system that measures the load that is exerted to maintain this physical state and static assessment of variations in temperature of the specimen, including in this part of the load cell, thermocouples and electronic structure with a / D converter, microcontroller and other components.

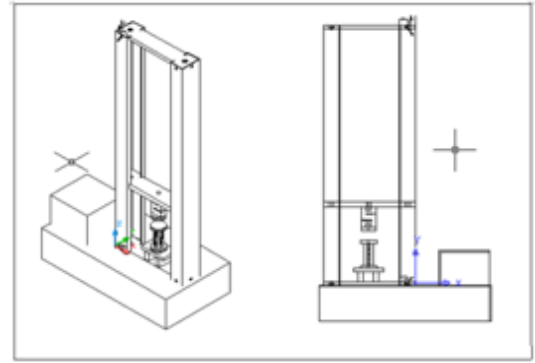

Figure 1 – Design of mechanical structure

The modular structure of the equipment is also innovative in order to allow a "personalization" of the tests, with the addition of new equipment, which are subject to a pre-programmed operation. For example, in the case of an interest in evaluating the effect of an electrical discharge or a controlled increase in temperature. This can be done automatically and controlled by the central equipment.

In this research, it adopted the application of a load cell model SB-2000 HB brand to two tons, three thermocouples type K-01 model MTK Minipa brand and generic 2K2 NTC thermistor as individual sensors. The structures of signal processing for each sensor are suitable for different electrical response of each one of them.

**2.1 Specimen**

The specimen was designed in order to withstand the high loads that will be submitted during the trial and facilitate cleaning and disassemble, thus facilitating the handling and avoiding waste for further analysis.

Another plus point was taken into account the need for sensors that were in contact with expansive mortar during the trial period, mainly temperature sensors. To this, the specimen was fitted with six lateral holes, three on each side spaced 2.5 cm with a diameter of 0.22 cm, allowing the inclusion of common thermocouples or other sensors similar diameters. Both the layout and the dimensions of the holes that were designed to provide an assessment along the height of the specimen, with a diameter as small as possible to avoid the sensors are subjected to excessive loads.

The complete specimen consists of three parts, as shown in Figure 2, a central part in the form of cylinder with a central hole an inch in diameter, with walls four inches thick steel, and two equal parts, circular, responsible for closing the two ends of this cylinder.

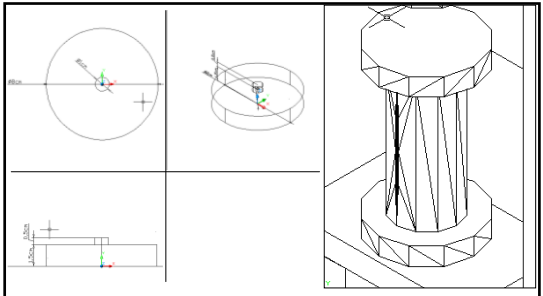

Figure 2 – Specimen Design

Parts of the specimen were performed in the garage Flee - Marble and Granite S / A. Dimensional tests were performed to measure the measures of the specimen, especially the side holes for the temperature sensors.

### 2.2 Data Acquisition System

The data acquisition system developed in individual blocks. Were the main food, part of the equipment viewing messages, using LCD display, microcontroller, keypad, filters and amplifier.

The connection of the acquisition system with its sensor conditioning circuits occurs from the J1 connector. Soon after this connector, we have analog filters responsible for cutting the Nyquist frequency of this circuit. After this filtering, the signal enters the system A / D conversion in the microcontroller ports RA0 to RA3. In the microcontroller, the signal is processed, the equipment shown on the display and sent to the door RD4 MAX232 chip, responsible for the adequacy of voltage CMOS microcontroller TTL serial port.

This data acquisition system has a resolution of 0.006 volts with a calibration range from 0 to 6 volts and thus corresponds to a change from 0 to 999 in value resulting from the conversion. The sampling rate is 200Hz. These data should guide the operations conditioning circuits responsible for processing signals from the sensors so that the unit change of the measured physical measurement corresponds to an increment or decrement the value obtained. For example, adding a pound on the load cell or a degree in thermocouple must correspond to a change of 0.006 volts in the signal conditioning.

The microcontroller selected for the initial design of the equipment was the PIC16F877A, shown in Figure 34. This choice was guided by the availability of resources of this component, among them we can highlight the A / D converter type SAR 10-bit resolution and the various communication possibilities that it presents (SPI, I²C, USART).

### 2.3 Operation of Equipment

The main feature of the operation of the equipment is its easy operation, thus simplified to the maximum the human-machine interface, reducing to a minimum the number of steps for the commencement of a trial.

Figure 3 shows the flowchart for use of equipment, where the main information available on the screen is the test time and the last charge read.

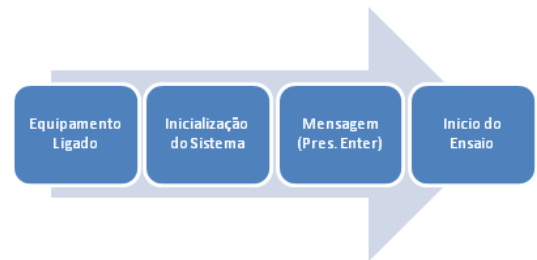

Figure 3 – Equipment usage

The limits for use in a continuous test is of 999 hours or 99,999 readings, thus giving a wide enough margin to analysis times of the load generated by the mortar.

### 2.4 Testing

The first step is the preparation of the test mass to be evaluated, which is performed according to the manufacturer's instructions.

After the preparation of the folder, you should proceed with the transfer to the metal mold, then close the mold, you clean the grout that will eventually be in excess and place the metal mold, pre-cleaned and lubricated, the testing machine itself scheduled for periodic readings of a total test time of 24 hours.

During the transfer to the mold, the process should be conducted in order to achieve the greatest possible uniformity of the mortar inside, avoiding the formation of air bubbles, voids or other substances.

In environmental terms, the test should be performed in a temperature (25 + 4) ° C with a relative humidity of (60 + 5)%. These conditions aim to reflect the normal weather conditions for most of Brazil.

Moisture is a critical factor due to high reactivity of calcium oxide with water, should be kept closed expansive mortar to the time of the test. After weighing, must immediately seal the container or package from which it was sampled to avoid hydration reactions that could interfere with subsequent testing.

Temperature has an important role in the kinetics of hydration reaction. Whereas most vendors have a specific composition for each temperature range, indicates that the range contains the value of the chosen standard temperature test.

## 3. Resultados de Discussão

For the test mortar business, we used an expansive mortar of Chinese origin and one of Italian origin, both provided by FUJI - Marble and Granite S / A, Unknown brand.

The tests were run on a 24 hour period, the Materials Characterization Laboratory, Department of Materials Engineering of UFCG, and the data captured on a local workstation and later analyzed. The result is shown in Figure 4.

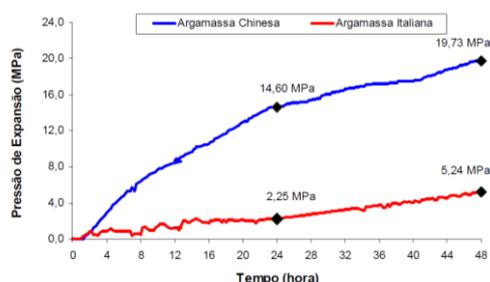

Figure 4 – Results from tested mortars

The mortar business is a relatively slow expansion, with a trend of growth rate. The expansive maximum pressure reached was 19.73 MPa to 5.24 MPa mortar and Chinese to Italian.

The mortar shows a performance of Chinese proprietary technology analyzed the expansion pressure of about 6.5 times higher than the Italian within 24 hours and this difference decreased to 3.8 times in 48 hours. Although both are widely used in the dismantling of ornamental when applied in situ (field), given the technical demands of the consumer market (miners), which indicates that the pressure for expansion of these mortars is higher than the resistance of ornamental rocks under practical conditions the extraction of ornamental rocks.

Curves outward pressure by the time it appears that the Chinese and Italian mortar expansion effectively begins after approximately 1 hour and 2 hours from baseline, respectively, suggesting a slower reaction of the mortar italiana.Com these results, observed that the pressure for expansion of Chinese and Italian plaster was only 21% and 3% from those reported by manufacturers (69 MPa). However, the value of the pressure of expanding commercial mortars determined by this method and using equipment developed Throughout this paper, will serve as a control parameter and comparison to the pressures of expansion obtained in mortar formulated and experimentally tested with the same methodology and held the same equipment for these mortar business.

Figure 5 presents the mean values of two pressure tests on six mortar expansion preliminarily formulated using the three types of retarders (sugar, plaster and retarding industrial base CMC) and varying concentrations of CaO, CaCO3 and cement. The test was run on a 24 hour period, for the practical application of these mortars under exploitation of ornamental stones this is the maximum acceptable for the dismantling of the rock occurs.

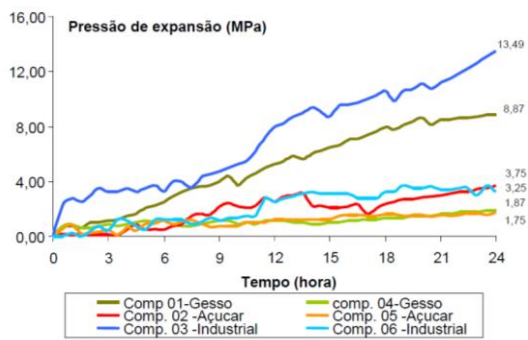

Figure 5 - Results from tested laboratorial mortars

From these preliminary formulations were able to get an overview of the relationship of pressure with different growth retardants of the hydration reaction of CaO, which is the physical-chemical principle that promotes the expansion of mixtures are used as expansive mortar applied to the dismantling of rocks ornamentais.O retardant industrial base CMC showed the best performance when compared to using sugar and plaster, for this same purpose. Moreover, because it is an industrial product compositions and quality control allow a reliability and reproducibility of their physical and chemical properties, allowing thus greater precision in different mixtures of mortars tested in this experimental work.

## 5. Conclusions

The objective of this work was the development and validation of an assay to test pressure rating of expansion through an expansive mortar equipment developed for it using development tools based on methodologies CAD, microcontrollers, and open source technology.

With the use of these tools and the adoption of the proposed methodology, it was possible to run considerable improvements in equipment pressure rating of expansive, consisting of the mechanical structure, circuit electronic data acquisition and test the computer program to analyze the results.

The new electronic structure met the initial expectations, and this stability and functionality demanded by the test methodology. The new electronic circuit for data acquisition was effective in converting the data obtained by the load cell into digital data be manipulated, either directly through the new LCD equipment, and indirectly by sending the information to PC via a serial communications port.

The coupling capacity of new sensors such as temperature or conductivity was available, both in the body of evidence through the side holes, as in the electronic circuit through additional data channels.

Being completed the improvements to the equipment, the next steps include the validation of the proposed methodology for testing by performing additional tests with mortars, both commercial and laboratory evaluation including temperature. These assessments should be performed in a controlled and appropriate to effect the qualification test.

.

**References**

[1] CAIMEX. Produtos Kayati SL-CRAS. www.caimex.com.br. Acesso em outubro de 2010.

[2] HUYNH, M.; LAEFER, D. F. Expansive cements and soundless chemical demolition agents : state of technology review. In: 11th Conference on Science and Technology, 2009, Ho Chi Minh. Anais... Ho Chi Minh: Vietnam Academy Of Science And Technology, Vietnam, 2009. Disponível em <http://irserver.ucd.ie/dspace/bitstream/10197/2285/1/79..pdf>. Acesso em 02 de novembro de 2010.

[3] HANIF, M., AL-MAGHRABI, M. N. H. Effective Use of Expansive Cement for the Deformation…. G.U. Journal of Science, Jeddah – Arabia Saudita, p. 1-5, 22 de novembro de 2006. Disponível em < http://www.fbe.gazi.edu.tr/dergi/ tr/ dergi/tam/20(1)/1.pdf >. Acesso em 7 de maio de 2007.

[4] SHIRO ISHI. Study of a Demolition Method Using Non Explosive Demolition Agent, R & D Laboratório de Novos Produtos, Onada Cement Co., Japão, 2006.

[5] KAWANO, et al. Patente nº 4,316,583. Disponível em <http://www.patft.uspto.gov>. Acesso em fevereiro de 2005.

[6] MOYER, Jr, et al. Patente nº 4,205,994. Disponível em <http://www.patft.uspto.gov>. Acesso em fevereiro de 2005.